\documentclass{article}
\usepackage{emulateapj,apjfonts}
\input psfig

\def\kms{km s$^{-1}$}

\def\mh{M_{\bullet}}
\def\ms{\mh-\sigma}
\def\msun{M_{\odot}}
\def\fun#1#2{\lower3.6pt\vbox{\baselineskip0pt\lineskip.9pt
  \ialign{$\mathsurround=0pt#1\hfil##\hfil$\crcr#2\crcr\sim\crcr}}}
\def\lap{\mathrel{\mathpalette\fun <}}

\lefthead{Merritt \& Ferrarese}
\righthead{$\mh-\sigma$ Relation}

\begin{document}

\title{The $\mh-\sigma$ Relation for Supermassive Black Holes}

\author{David Merritt and Laura Ferrarese}
\affil{Rutgers University, Dept. of Physics and Astronomy, New Brunswick, NJ, 08854}
\authoraddr{Department of Physics and Astronomy, 136 Frelinghuysen Road, 
Piscataway, NJ 08854}

\begin{abstract}
We investigate the differences in the $\ms$ relation derived recently
by Ferrarese \& Merritt and by Gebhardt and collaborators.  The
shallower slope found by the latter authors ($3.75$ vs. $4.8$) is due
partly to the use of a regression  algorithm that ignores measurement
errors, and partly to the value of the velocity dispersion adopted for
a single galaxy, the Milky Way.  A steeper relation is shown to
provide a better fit to black hole masses derived from reverberation
mapping studies.  Combining the stellar dynamical, gas dynamical, and
reverberation  mapping mass estimates, we derive a best-fit relation
$\mh = 1.30 (\pm 0.36) \times 10^8\msun(\sigma_c/200\ {\rm km\
s}^{-1})^{4.72(\pm 0.36)}$.
\end{abstract}


\section{Introduction}

Ferrarese \& Merritt (2000; FM00; Paper I) demonstrated a tight
correlation  between the masses of supermassive black holes (BHs) and
the velocity  dispersions of their host bulges,
$\mh\propto\sigma^{\alpha}$,  $\alpha=4.8\pm 0.5$.  The scatter in the
relation was found to be consistent with that expected on the  basis
of measurement errors alone; in other words, the underlying
correlation between $\sigma$ and $\mh$ is essentially perfect.  The
relation is apparently so tight that it surpasses in predictive
accuracy what can be achieved from detailed dynamical modelling of
stellar  kinematical data in most galaxies.  As an example, FM00
showed that the BH mass estimates of  Magorrian et al. (1998), derived
from ground-based optical observations, lie systematically above the
$\ms$ relation defined by galaxies with secure BH masses, some by as
much as two orders of magnitude.

The $\ms$ relation of Paper I was based on central velocity
dispersions $\sigma_c$, corrected to an effective aperture of radius
$r_e/8$, with $r_e$ the half-light radius.  Central velocity
dispersions are easily measured and available for  a large number of
galaxies (Prugniel et al. 1997).  An alternative form of the $\ms$
relation was investigated by  Gebhardt et al. (2000a; G00) who used
$\sigma_e$ as the independent variable; $\sigma_e$ was defined as the
spatially-averaged, rms, line-of-sight stellar velocity within the
effective radius $r_e$.  Computing $\sigma_e$ requires knowledge of
the stellar rotation and velocity dispersion profiles at all radii
within $r_e$, as well as information about the inclination of the
rotation axis with respect to the line of sight.  These data are
available for a smaller number of galaxies; on the other hand,
$\sigma_e$ might be expected to reflect the depth of the stellar
potential well more accurately than $\sigma_c$.

The versions of the $\mh-\sigma$ relation derived by  FM00 and by G00
differ in two important ways.  The latter authors found a
significantly smaller slope ($\alpha=3.75\pm 0.3$ vs. $4.8\pm 0.5$) as
well as a greater vertical scatter -- greater  both in an absolute
sense,  and relative to measurement errors in $\mh$.  G00 estimated
that approximately $40\%$ of the scatter in $\mh$  about the mean line
was intrinsic and the remainder due to measurement errors.  FM00 found
no evidence for an intrinsic scatter in $\mh$.

The $\ms$ relation is currently our best guide to BH demographics, and
it is important to understand the source of these differences.  That
is the goal of this paper.  In addition to using different measures
of the velocity dispersion, FM00 and G00 analyzed different galaxy
samples,  and used different algorithms for fitting regression lines
to the data.  We find that regression algorithms that account
correctly for errors in the measured variables always give a steeper
slope  than that found by G00.  We also show that the steeper relation
derived by FM00 provides a better fit to galaxies with BH masses
computed by reverberation  mapping.

\section{Data}

Table 1 gives the data used here.  The first 12 galaxies (Sample 1)
are ``Sample A'' from Paper I, consisting of those galaxies with
published BH mass estimates that were deemed reliable -- roughly
speaking, galaxies in which the sphere of influence of the BH has been
resolved.  Five of these masses are derived from stellar kinematics
and seven  from gas dynamics.  All of these galaxies were included in
the G00 sample as well, with the exception of NGC 3115; for this
galaxy, we assume $\sigma_e=\sigma_c$.   The second part of Table 1
contains the additional 15 galaxies  included by G00 (Sample 2).$^1$
\footnotetext[1]{The error bars plotted in Fig. 2 of G00 do not always
correspond to the values listed in their Table 1 (e.g. NGC 4291 and
NGC 5845). We used the tabulated values.}  Most of the BH mass
estimates for these galaxies are based on unpublished STIS data.  In
addition, G00 included M31 and NGC 1068,  which were excluded from
Paper I on the grounds that their BH masses  were deemed unreliable.
We computed distances for the G00 galaxies in Table 1 in the same way
as  in Paper I and corrected the BH masses accordingly.  We also
computed aperture-corrected central dispersions $\sigma_c$ for the G00
galaxies.

At this point, we are already in a position to test the idea, proposed
by G00, that the steeper slope of the $\mh-\sigma_c$ relation in Paper
I is due to spuriously high values of $\sigma_c$ for the more nearby
galaxies. This idea is rejected based on Figure 1, which shows that
there is remarkably little difference on average between $\sigma_c$
and $\sigma_e$.  This is presumably due to the flatness of galaxy
rotation and velocity dispersion profiles, and to the fact that even
$\sigma_c$ is measured on a large enough scale that it is essentially
unaffected by the presence of the BH.  The mean ratio of $\sigma_e$ to
$\sigma_c$ is $1.01$; the correlation coefficient of $\log\sigma_e$ vs
$\log\sigma_c$  is $0.97$.

\hskip -2.55in\psfig{file=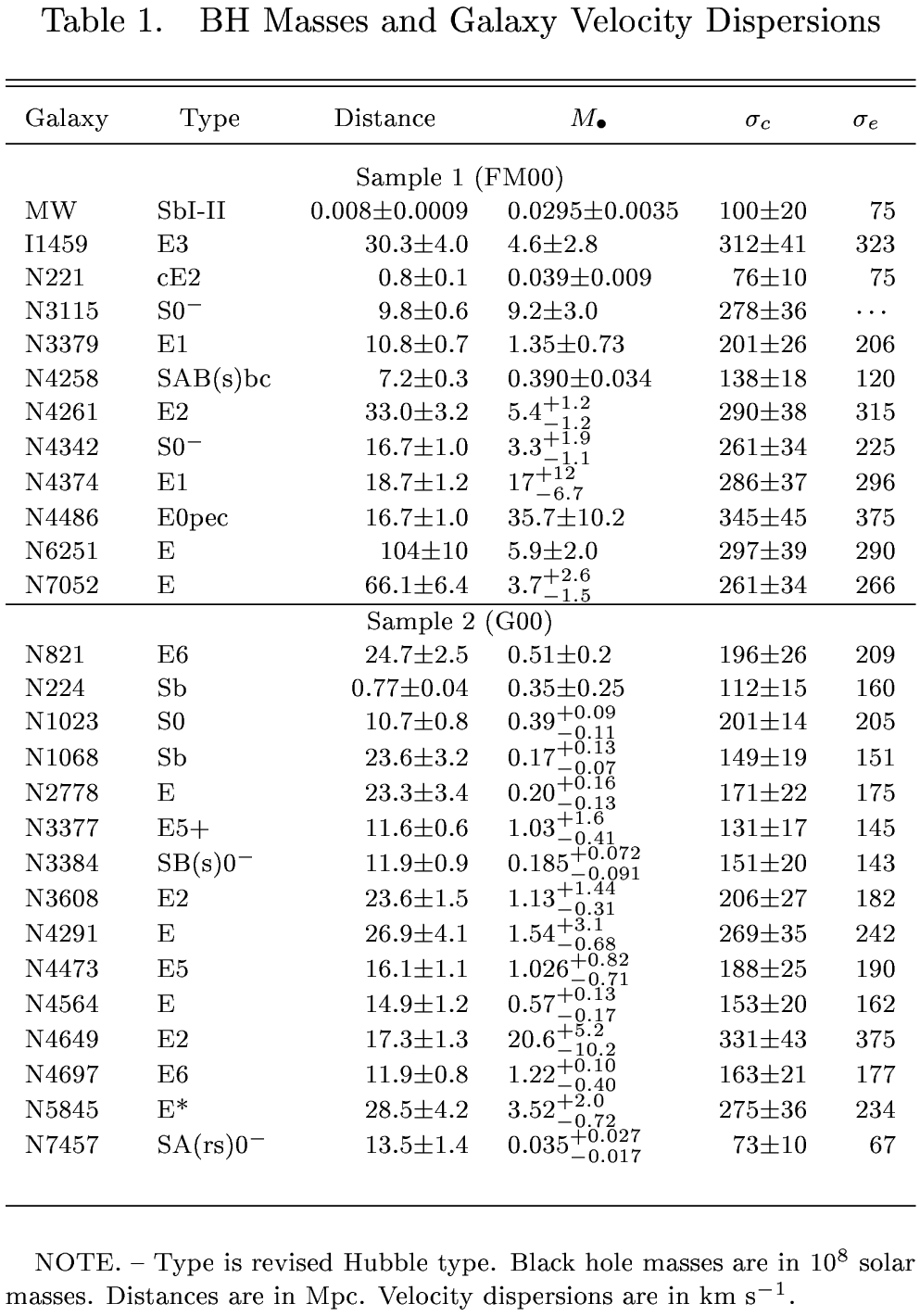,width=18.0cm,angle=0}
\vskip -4.3in

However, we notice that $\sigma_e$ and $\sigma_c$ differ significantly
for one particular galaxy, the Milky Way. G00 adopted a value of
$\sigma_e =$ 75 \kms for the Galaxy based on the velocity dispersion
between 50 and 500 arcsec (Kent 1992; Genzel et al. 2000).  They
apparently neglected to account for the contribution of the rotational
velocity, which in the same region is 103$\pm$15 \kms (Kent 1992).
More importantly, 500 arcsec corresponds to a projected radius of 20
pc at the Galactic center, more than two orders of magnitude smaller
than the effective radius of the Galactic bulge ($\sim 2.7$ kpc;
cf. Gilmore, King \& van der Kruit 1990).  The bulge velocity
dispersion has been measured by several authors at various
Galactocentric distances within 4 kpc, all giving values between 75
and 110 \kms, with a tendency for $\sigma$ to increase slowly toward
the center (e.g. Kent 1992; Tiede \& Terndrup 1997, 1999; Minniti
1996; C\^ot\'e 1999; Zhao et al. 1996).  The rotational velocity in
the inner 1.5 kpc is well approximated by a solid body curve with $v
\sim 65 - 87$ \kms kpc$^{-1}$ (e.g.  Tiede \& Terndrup 1997, 1999;
Morrison \& Harding 1993; Menzies 1990; Kinman, Feast \& Lasker 1988).
In view of these results, we question the choice of $\sigma_e = 75$
\kms for the Milky Way.  FM00 adopted $\sigma_c = 100$ \kms; in what
follows we will perform regression analyses assuming values of both 75
\kms and 100 \kms for $\sigma_e$.  We will show that the slope of the
$\mh-\sigma_e$ relation depends significantly on which value is used.

G00 assumed constant errors on log$\mh$ and zero measurement errors in
$\sigma_e$ when carrying out their least-square fits. However,
ignoring measurement errors in the independent variable is well known
to bias the slope downwards (e.g Jefferys 1980).  Even when high
signal-to-noise data are used, measurement errors on the velocity
dispersions are easily at the 10\% level (e.g. 

\psfig{file=figure1.epsi,width=8.8cm,angle=0} \figcaption{Comparison
of $\sigma_c$ (the central velocity dispersion) and $\sigma_e$ (the
rms  velocity within one effective radius)  for the galaxies in Table
1. The solid line has a slope of one.}
\vskip 0.3in

\noindent van der Marel et
al. 1994), and cannot be neglected. Unfortunately, the data used by
G00 to compute $\sigma_e$ are mostly unpublished, and the authors do
not give error estimates in their paper.  Therefore, in the regression
analyses described below we will make various assumptions about the
measurements uncertainties in $\sigma_e$.

To understand how the different galaxy samples used by FM00 and by G00
may have affected their respective conclusions about the $\ms$
relation, we analyze Sample 1  (the 12 galaxies from Paper I) and
Sample 2 (the additional 15 galaxies  from G00) separately.  The BH
masses in Sample 2 are significantly less accurate than those in
Sample 1, with an rms uncertainty of  $0.28$ dex, compared to $0.18$
dex for Sample 1.  We also present results from the analysis of the
entire  set of 27 galaxies, called the ``combined sample'' below.

\section{Analysis}

We assume a relation of the form
\begin{equation}
Y_i = \alpha X_i + \beta + \epsilon_i
\end{equation}
between the measured variables, where $Y$ is $\log\mh$ and $X$ is
either $\log\sigma_c$ or $\log\sigma_e$.  Units are solar masses for
$\mh$ and \kms for $\sigma$.  The $\epsilon_i$ describe measurement
errors as well as intrinsic scatter in the relation, if any.  A large
number of regression algorithms are available for recovering estimates
$\hat\alpha$ and $\hat\beta$ of the slope and intercept and their
uncertainties, given ($X_i$, $Y_i$) and their estimated errors
($\sigma_{Xi},\sigma_{Yi}$).  These algorithms differ in the degree of
generality of the model that is  assumed to underly the data.  The
following four algorithms were used here.

{\it Ordinary Least-Squares} (OLS). All of the error is assumed to lie
in the dependent variable (i.e. $\log\mh$) and the amplitude of the
error is assumed to be the same from measurement to measurement.  This
is the algorithm adopted by G00.  We use the implementation {\tt
G02CAF} from the NAG subroutine library. The OLS estimator is biased
if there are measurement errors 

\hskip -4.7in\psfig{file=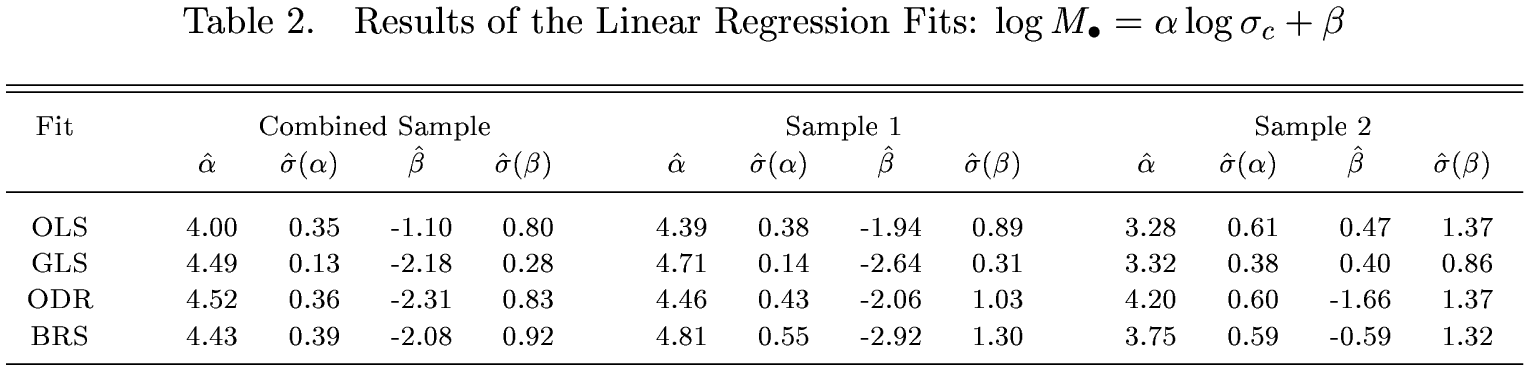,width=17.5cm,angle=0}
\vskip -3.0in

\noindent in the independent variable, or if the errors in the
dependent variable vary from data point to data point (e.g. Jefferys
1980).

{\it General Least-Squares} (GLS). All of the error is still assumed
to reside in the dependent variable, but the amplitude of the error
may vary from point to point.  Press et al. (1989) implement this
model in their routine {\tt fit}, which we use here.

{\it Orthogonal Distance Regression} (ODR). The underlying variables
are assumed to lie exactly on a straight line, i.e. to have no
intrinsic scatter, but the observed quantities are allowed to have
measurement errors in both $X$ and $Y$, which may differ from point to
point. This model is incorporated in the routines {\tt fitexy} of
Press  et al. (1989) and {\tt fv} of Fasano \& Vio (1988).  We use the
former routine here; the latter was found to give essentially
identical results. The ODR estimator may be biased if the true
variables exhibit intrinsic scatter about the linear relation, in
addition to measurement errors (e.g. Feigelson \& Babu 1992).

{\it Regression with Bivariate Errors and Intrinsic Scatter} (BRS).
As in ODR, the data are permitted to have measurement errors in both
$X$ and $Y$ that differ from point to point.  In addition, the
underlying variables are allowed to have an intrinsic scatter about
the regression line. We use the routine BCES($Y|X$) of Akritas \&
Bershady (1996), the same routine used in Paper I.

\hskip -0.6in\psfig{file=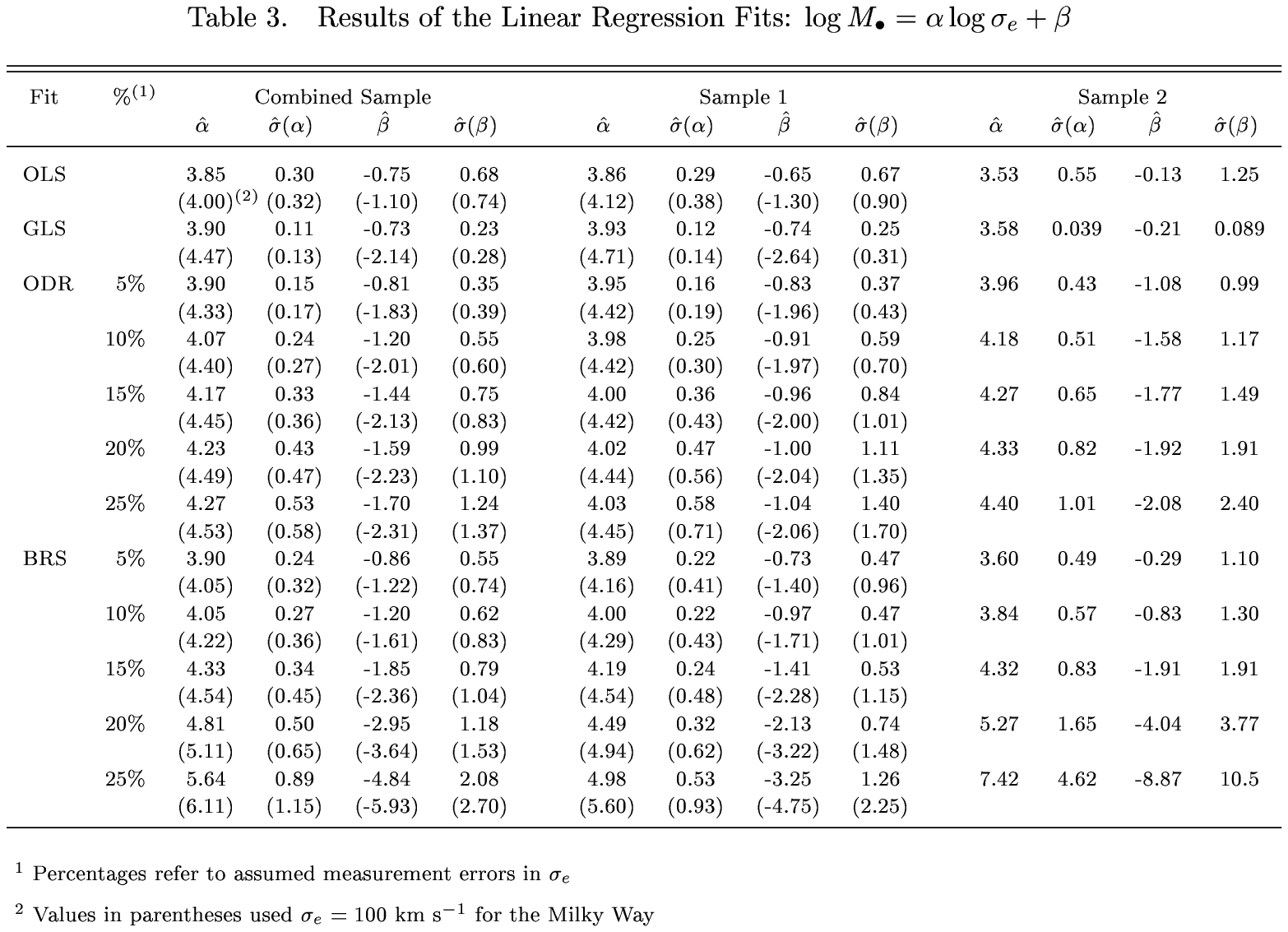,width=17.5cm,angle=0}
\vskip -3.0in

Tables 2 and 3 give estimates of the slope and intercept, $\hat\alpha$
and $\hat\beta$, and their uncertainties as computed by each of  the
four algorithms, using $\sigma_c$ and $\sigma_e$ as independent
variables.  Values in parentheses correspond to setting $\sigma_e=100$
\kms for the Milky Way, as discussed above.  The results are
summarized below and in Figure 2.

1. Accounting for errors in one or both variables increases the slope
of the relation, whether expressed in terms of $\sigma_c$ or
$\sigma_e$.  Ignoring measurement errors biases the slope too low, for
two reasons.  The BH masses in Sample 2 are significantly more
uncertain than those in Sample 1 and, as a group, exhibit a shallower
slope (particularly when expressed in terms of $\sigma_c$); routines
like OLS that weight all data equally therefore underestimate the true
slope.  Second, ignoring measurement errors in the independent
variable  ($\log\sigma$) {\it always} yields spuriously low slopes
(e.g. Jefferys 1980).  We find that the shallowest slope for every
sample  is returned by OLS, the routine used by G00.  All other
algorithms give slopes  in the range $4\lap\hat\alpha\lap 5$ for the
combined sample.

2. The slope inferred for the $\mh-\sigma_e$ relation using ODR and
BRS depends somewhat on the assumed errors  in $\sigma_e$.  Increasing
the assumed error from $5\%$ to $20\%$  increases the BRS slope of the
combined sample  from $3.9$ to $4.8$.

\newpage

\vspace*{-2.95in}
\hskip -0.5in \psfig{file=figure2.epsi,width=20.8cm,angle=0}
\vskip -2.0in

3. Even when the appropriate fitting routines are used, the
$\mh-\sigma_c$ relation tends to have a steeper slope than the
$\mh-\sigma_e$ relation. This difference, however, is driven by one
galaxy only: when $\sigma_e$ for the Milky Way is increased from 75
\kms~(used by G00) to a more appropriate -- in our opinion -- value of
100 \kms, both relations have a best-fit slope of $\sim 4.5\pm
0.5$ for the combined sample (assuming a plausible 10\% - 15\% error
on $\sigma_e$).

4. Adding the galaxies of Sample 2 (from G00) has little impact on the
results, as long as measurement errors are taken into account by the
fitting routine and $\sigma_e = \sigma_c = 100$ \kms~is used for the
Galaxy: the regression lines for Sample 1 (from Paper I) and for the
combined sample are essentially the same.  In other words, the BH
masses added by G00 are too uncertain to significantly alter the fit
determined by the galaxies from Sample 1 alone.

We conclude that the different slopes found by G00 and FM00 ($3.75$ vs
$4.8$) are due partly to the neglect of measurement errors by the
former authors, and partly to the difference between $\sigma_e$ and
$\sigma_c$ for a  single data point, the Milky Way.  If we use the
more appropriate value of $\sigma_e=100$ \kms for the Milky Way, and a
plausible 10\% - 15\% error on $\sigma_e$, the $\mh-\sigma_c$ and
$\mh-\sigma_e$ relations have essentially the same slope, $\sim 4.5$.
The data points added by G00,  based mostly on unpublished

\noindent 
\vspace*{5.625in}
~~~~~~~~~~~~~~~~~~~~~~~~~~~~~~~~~~

\noindent modelling of stellar kinematical data from STIS, appear to contain 
little information about the $\ms$ relation that was not already 
contained in the more accurate masses from  Paper I.

We next address the scatter in the $\ms$ relation.
The $\chi^2$ merit function for a linear fit to data with errors
in both variables is 
\begin{equation}
\tilde\chi^2 = {1\over N-2}\sum_{i=1}^N 
{\left(Y_i-\hat\alpha X_i-\hat\beta_i\right)^2\over 
\sigma_{Y,i}^2 + \hat\alpha^2\sigma_{X,i}^2}
\end{equation}
(e.g. Press et al. 1989),
where, in our case, $Y=\log\mh$ and $X=\log\sigma$.
A good fit has $\tilde\chi^2\lap 1$.
Since measurement uncertainties $\sigma_{X,i}$ are not available
for the $\sigma_e$,
we computed $\tilde\chi^2$ only for the $\mh-\sigma_c$ relation.
We are also interested in the absolute scatter in $\log\mh$,
which we define as 
\begin{equation}
\Delta_{\bullet} = \sqrt{{1\over N}\sum_{i=1}^N\left(Y_i-\hat\alpha X_i
-\hat\beta_i\right)^2}.
\end{equation}
We computed $\tilde\chi^2$ and $\Delta_{\bullet}$ using the 
fits given by the BRS regression algorithm (Table 2).
The results are:

\setcounter{figure}{2}
\psfig{file=figure3.epsi,width=8.8cm,angle=0}
\figcaption{Reverberation mapping masses for seven galaxies.
Solid line is the $\mh-\sigma_c$ relation from 
Ferrarese \& Merritt (2000);
dashed line is the $\mh-\sigma_e$ relation from 
Gebhardt et al. (2000a).}
\vskip 0.3in 

\begin{eqnarray}
{\rm Sample\ 1:}\ \  \tilde\chi^2 & = & 0.74 \ \ \ \ \ \ \Delta_{\bullet}=0.26
\nonumber \\
{\rm Sample\ 2:}\ \  \tilde\chi^2 & = & 1.67 \ \ \ \ \ \ \Delta_{\bullet}=0.35
\nonumber \\
{\rm Combined\ sample:}\ \  \tilde\chi^2 & = & 1.20 \ \ \ \ \ \ \Delta_{\bullet}=0.34
\nonumber
\end{eqnarray}
The Sample 2 galaxies exhibit a larger scatter in $\log\mh$ than the
galaxies in Sample 1 ($0.35$ dex vs. $0.26$ dex), consistent with
their greater measurement uncertainties (Table 1).  Furthermore the 12
galaxies from Paper I define a significantly  tighter correlation, as
measured by $\tilde\chi^2$,  than the 15 galaxies added by G00
($\tilde\chi^2=0.74$ vs. $\tilde\chi^2=1.67$),  or than the combined
sample.  Thus we confirm the conclusion of G00 that the scatter in
their data about the best-fit linear relation exceeds that expected on
the basis of measurement error alone.  The large $\tilde\chi^2$ for
Sample 2 may indicate that the measurement uncertainties quoted by G00
are too small.

\section{Reverberation Mapping Masses}

A long-standing discrepancy exists between BH masses determined from
stellar kinematics and from reverberation mapping; the latter
technique uses emission lines in active galactic nuclei (AGN) to probe
the virial mass within the broad-line region (Netzer \& Peterson
1997).  Since there are currently no galaxies with BH masses
determined independently by the two techniques, any comparison must be
statistical.  The standard approach (e.g. Wandel 1999) has been to
compare the average BH mass at a given bulge luminosity as computed
from reverberation mapping with the mass predicted by the Magorrian et
al.  (1998) relation; the latter is based on stellar kinematical data,
mostly of low spatial resolution.  The discrepancy is a factor of
$\sim 20$ in the sense that the  reverberation-mapping masses are too
low (Wandel 1999).  This discrepancy has most often been attributed to
some unspecified, systematic error in the reverberation mapping masses
(e.g. Richstone et al. 1998; Faber 1999; Ho 1999).

FM00 showed that the Magorrian et al. masses fall systematically above
the $\mh-\sigma_c$ relation defined by galaxies with more secure BH
mass estimates, some by as much as two orders of magnitude.  The
offset is strongly correlated with distance suggesting a systematic,
resolution-dependent error in the Magorrian et al.  modelling.  Much
or all of the discrepancy with the reverberation mapping masses might
therefore be due to systematic errors in the  Magorrian et al. masses,
contrary to the usual assumption.  Gebhardt et al. (2000b) tested this
idea by plotting seven AGN BH masses against their $\mh-\sigma_e$
relation.$^2$ \footnotetext[2]{The velocity dispersions plotted by
Gebhardt et al.  (2000b) are labelled $\sigma_e$ even though they are
central values.}  We reproduce that plot here, as Figure 3.  The fit
is reasonable, although the points tend to scatter  below the line.
We also plot in Figure 3 the steeper $\mh-\sigma_c$ relation  derived
in Paper I (given here, in Table 2,  as the BRS regression fit on
$\sigma_c$ for Sample 1).  The steeper relation of Paper I is clearly
a better fit.

We stress that several of the AGN data points lie at the low-mass end
of the distribution where the $\ms$ relation is strongly affected by
uncertainties in the slope.  Nevertheless, there would no longer
appear to be any {\it prima facie} reason for believing that the
reverberation mapping  masses are systematically in error.
Furthermore the scatter in these masses about the $\ms$ relation
appears to be comparable to that of the Sample 2 data from G00.  We
therefore carried out regression fits combining the reverberation
mapping masses with Sample 1 and Sample 2, for a total of 34 galaxies.
We assumed $50\%$ measurement errors in the AGN $\mh$ and $15\%$
errors in the $\sigma_c$.  The results (using the BRS regression
routine) were:
\begin{equation}
\hat\alpha = 4.72 \pm 0.36,  \ \ \ \ \ \ \hat\beta = -2.75 \pm 0.82,
\end{equation}
very close to the parameters derived in Paper I using Sample 1 alone.
This fit has $\tilde\chi^2 = 1.11$ and $\Delta_{\bullet}=0.35$, about
as good as obtained using the galaxies in Table 1.

\section{Summary}

We investigated the differences in the $\ms$ relation as derived by
Ferrarese \& Merritt (2000) and by Gebhardt et al. (2000a).  The
latter authors found a shallower slope ($3.75$ vs. $4.8$) and a
greater vertical scatter, larger than expected on the basis of
measurement errors alone.  Three possible explanations for the
differences were explored: different galaxy samples; different
definitions of the velocity dispersion, central ($\sigma_c$) vs
integrated ($\sigma_e$); and different routines for carrying out the
regression.  The shallower slope of the G00 relation was found to be
due partly to the use  of a regression algorithm that does not account
properly for  measurement errors, and partly to the adoption of a
value of $\sigma_e$  for the Milky Way which is, in our opinion,
implausibly low.  The greater scatter seen by G00 is due to larger
uncertainties associated with the additional BH masses included by
them, mostly from unpublished STIS data.  When measurement
uncertainties are properly accounted for, the parameters of the
best-fit relation derived from the  combined samples of FM00 and G00
are essentially identical to  those derived from the sample of FM00
alone.  The steeper relation derived by FM00 also provides  a better
fit to BH masses obtained from reverberation mapping.  A regression
fit to the combined sample of 34 galaxies,  including stellar
dynamical, gas dynamical, and reverberation  mapping masses, yields:

\begin{equation}
\mh = 1.30 (\pm 0.36) \times 10^8\msun\left({\sigma_c\over 200\ {\rm
km\ s}^{-1}}\right)^{4.72(\pm 0.36)}.
\end{equation}

The scientific implications of Equation (4) are discussed briefly
by FM00, and extensively in Merritt \& Ferrarese (2000).
This relation is essentially identical to the one derived in Paper I.
We suggest that there is no longer any reason to assume, as a number
of authors (Richstone et al. 1998; Faber 1999; Ho 1999) have done,
that the reverberation mapping masses are less accurate than masses
derived from stellar kinematics (Magorrian et al. 1998).

We stress that the current sample of galaxies with reliable BH mass
estimates is likely affected by severe selection biases, which are
very difficult to quantify. Our results highlight the need for
accurate BH masses if the $\ms$ relation is to be further refined.
Only a handful of galaxies observed with STIS are likely to yield mass
estimates as accurate as those already available  for the galaxies in
Paper I.  Uncertainties in the reverberation mapping masses are
probably comparable to those obtained from HST data in most galaxies;
however the number of galaxies with reverberation mapping masses is
large ($\sim 35$) and growing.  Furthermore, many of these galaxies
are in the critical, low mass range, $10^6\lap\mh\lap 10^8 \msun$.  An
aggressive campaign to measure stellar velocity dispersions in AGN
might be the best route toward refining the $\ms$ relation.

\bigskip
\acknowledgments

LF acknowledges grant  NASA NAG5-8693, and DM acknowledges grants NSF
AST 96-17088 and NASA NAG5-6037.   We thank Brad Peterson for useful
discussions.

\end{document}